\iffalse Before sending further, pass this check list:
  [ ] find and resolve all "??" and "\ifnote"
  [ ] Spell-check
  [ ] order of using/defining abbreviations
  [ ] citation order (use RefTest)
  [ ] figure order & reference order
  [ ] Check LaTeX  output files (*.log) for warnings
  [ ] Check BibTeX output (screen) for warnings
  [ ] update PACS
  [ ] decide on color figures
  [ ] Re-read paper in the morning

After completing this list, if you made at least one correction,
re-do this check-list from the beginning, until no corrections will
be done.\fi

\documentclass[twocolumn,amsmath,amssymb,aps,preprintnumbers,showpacs,superscriptaddress]{revtex4}

\usepackage{graphicx}
\usepackage{dcolumn}
\usepackage{bm}
\usepackage[dvips]{color} 

\newcommand{\lapprox}{{\scriptscriptstyle\stackrel{<}{\sim}}}

\newcommand{\co}[2]{\ifcase #1 \or #2 \fi}

\begin{document}



\title{Suppression of dissipation in Nb thin films with triangular
antidot arrays by random removal of pinning sites}

\author{M.~Kemmler}\email{kemmler@pit.physik.uni-tuebingen.de}
\author{D.~Bothner}%
\affiliation{%
Physikalisches Institut -- Experimentalphysik II and Center for
Collective Quantum Phenomena and their Applications, Universit\"{a}t
T\"{u}bingen, Auf der Morgenstelle 14, D-72076 T\"{u}bingen, Germany
}%
\author{K.~Ilin}
\author{M.~Siegel}
\affiliation{ IMS, Universit\"{a}t Karlsruhe, Hertzstr.~16, D-76187
Karlsruhe, Germany
}%
\author{R.~Kleiner}
\author{D.~Koelle}
\affiliation{%
Physikalisches Institut -- Experimentalphysik II and Center for
Collective Quantum Phenomena and their Applications, Universit\"{a}t
T\"{u}bingen, Auf der Morgenstelle 14, D-72076 T\"{u}bingen, Germany
}%

\date{\today}

\begin{abstract}
The depinning current $I_c$ versus applied magnetic field $B$ close
to the transition temperature $T_c$ of Nb thin films with randomly
diluted triangular arrays of antidots is investigated.
Our experiments confirm essential features in $I_c(B)$ as predicted
by Reichhardt and Olson Reichhardt [Phys.~Rev.~B {\bf 76}, 094512
(2007)].
We show that, by introducing disorder into periodic pinning arrays,
$I_c$ can be enhanced.
In particular, for arrays with fixed density $n_p$ of antidots, an
increase in dilution $P_d$ induces an increase in $I_c$ and decrease
of the flux-flow voltage for $B>B_p=n_p\Phi_0$.
%
\end{abstract}

\pacs{74.25.Qt, 74.25.Sv, 74.78.Na}
%
%
%
%



\maketitle


The investigation of vortices in type-II superconductors in the
presence of tailored pinning potential landscapes has attracted a lot
of theoretical and experimental interest.
On the one hand, vortices in superconductors may act as a model
system in order to investigate general properties such as the
dynamics and phase transitions in systems of interacting particles
(e.~g.~ colloidal suspensions \cite{Korda02,Pine05}, Wigner crystals
\cite{Andrei88} , charge density waves \cite{Gruner88} or various
types of ratchets and Brownian motors
\cite{Magnasco93,Reimann02,Haenggi05}).
On the other hand, the ability to manipulate and control the static
and dynamic properties of vortices is fundamental for superconducting
device applications \cite{Moshchalkov00}.

Modern lithography techniques allow the placement of artificial
pinning sites into superconducting thin films with well-defined
size, geometry and spatial arrangement.
In case of periodic arrangements, enhanced vortex pinning was found
for magnetic fields, at which the vortex lattice is commensurate with
the pinning array
\cite{Daldini74,Fiory78,Baert95,Martin97,VanBael99,Villegas03a}.
The enhanced pinning leads, e.g., to peaks in the critical depinning
current $I_c$ at multiples of a so-called first matching field
$B_p=n_p\Phi_0$; here the density of vortices carrying one flux
quantum $\Phi_0=h/2e$ equals the density of pinning sites $n_p$.
However, at non-matching fields the vortex lattice is less pinned
due to elastic deformations and formation of interstitial vortices.
Hence, the question arises whether other arrangements -- between the
two extremes of periodic and random pinning arrangements -- may lead
to an enhanced vortex pinning over a broader range of applied
magnetic field $B$.

Recently, it has been shown by numerical simulations
\cite{Misko05,Misko06a} and experimentally
\cite{Villegas06,Kemmler06,Silhanek06} that a quasiperiodic
arrangement of pinnig sites produces additional commensurability
effects and hence an enhanced pinning below the first matching field.
A different proposal was made very recently by Reichhardt and Olson
Reichhardt \cite{Reichhardt07a}.
By molecular dynamics simulations they investigated periodic pinning
arrays that have been diluted, by randomly removing the fraction
$P_d$ of pins, while keeping the pin density $n_p$ fixed.
Such arrays are very interesting, since with increasing dilution the
pinning potential undergoes a gradual transition from periodic to
purely random.
Therefore, this model is suitable to explore the intermediate region
between order and disorder, as it is usually found in real world.
Interestingly, the simulations showed that the introduction of some
disorder leads to an enhanced critical current above the first
matching field.
In periodic pinning arrays the vortices sitting at the pinning sites
form easy flow channels for interstitial vortices \cite{Velez02},
while for randomly diluted pinning arrays channeling should be
suppressed \cite{Reichhardt07a}.
This approach is also interesting from a general point of view, as
the presence of disorder in competition with periodic potentials is
also investigated in many other physical systems,
e.g.~two-dimensional conductors \cite{Dorn05}, Ising ferromagnets
\cite{Griffiths69}, and Josephson Junction arrays
\cite{Li91a,Granato01}.
For related recent work on vortex phases see \cite{Pogosov08}.

In this work, we present results on the experimental investigation of
vortex pinning and flow in superconducting Nb thin films containing
randomly diluted triangular arrays of submicron holes (antidots) as
pinning sites.
We studied $I_c(B)$ at variable temperature $T$ close to the
superconducting transition temperature $T_c$, and we compare pinning
arrays with different dilution, considering two different scenarios:
{\em (i) ''Scaled lattices``:}
For different values of $P_d$, we fix the density $n_p$ of pinning
sites.
Accordingly, the lattice parameter (smallest separation between
pinning sites) scales as $a(P_d)=a(0)\sqrt{1-P_d}$.
In this scenario a controlled transition from periodic to random
arrangement of antidots is investigated.
{\em (ii) ''Fixed lattices``:}
Here, we fix the lattice parameter $a$ for different values of
$P_d$.
Accordingly, the density of pinning sites scales as
$n_p(P_d)=(1-P_d)\,n_p(0)$.
Here, with increasing $P_d$ a transition to plain films (no antidots)
is treated.

Our experimental results confirm essential features as predicted in
\cite{Reichhardt07a}:
The fixed and scaled lattices show two different kind of matching
effects, which differently depend on temperature.
Furthermore, the scaled lattices show an enhancement of $I_c$ at
magnetic fields above $B_p$ with increasing dilution $P_d$.
This effect is caused by suppression of vortex channeling and can be
also observed in the dynamic regime, i.e.~by measuring
current-voltage ($IV$) characteristics.


The experiments were carried out on
$d=60\,$nm thick Nb films which were deposited by dc magnetron
sputtering in the same run on four separate Si substrates with
$1\,\mu$m thick $\rm SiO_2$ on top.
Patterning was performed by e-beam lithography and lift-off to
produce Nb bridges of width $W=200\,\mu$m and length $L=640\,\mu$m.
The bridges contain circular antidots (diameter $D=260\ldots
550\,$nm), arranged in a triangular lattice that has been randomly
diluted, with dilutions $\rm P_d$=0 (''undiluted array``), 0.2, 0.4,
0.6, 0.8, and 1 (''plain`` film, without antidots).
Each chip (\#1 to \#4) contains two or three sets (A, B, C) of
bridges.
Each set has six bridges with different values for $P_d=0\ldots 1$.
The antidot diameter $D$ is kept constant within each set and varies
from set to set.
The chips \#1 and \#2 contain sets of bridges with fixed lattice
parameter $a=1.5\,\mu$m.
I.~e.~the density of vertices of the corresponding triangular lattice
is $n_l=\frac{2}{\sqrt{3}}\frac{1}{a^2}=0.5\,\mu{\rm m}^{-2}$, which
corresponds the ''lattice matching field'' $B_l\equiv
n_l\Phi_0=1.1\,$mT (denoted as $B_\phi^*$ in \cite{Reichhardt07a}).
For those two chips, the antidot density $n_p$ decreases from $0.5$
to $0.1\,\mu{\rm m}^{-2}$ with increasing $P_d$ from 0 to 0.8.
The chips \#3 and \#4 contain sets of bridges with scaled lattice
parameters $a(P_d)=3.4\,\mu$m to $1.5\,\mu$m for $P_d=0$ to
$P_d=0.8$, respectively, in order to have a fixed antidot density
$n_p=0.1\,\mu\rm m^{-2}$ and ''pin density'' matching field
$B_p=0.21\,$mT (denoted as $B_\phi$ in \cite{Reichhardt07a}), with
$N_p\approx 12,500$ antidots in each bridge.
Below we present results obtained on bridges from sets \#1-B
($D=300\,$nm), \#3-B ($D=450\,$nm), 
\#4-B ($D=360\,$nm) and \#4-A ($D=260\,$nm).


To characterize our devices, we first measured resistance $R$ vs.~$T$
at $B=0$ and determined $T_c$ and normal resistance $R_n\equiv
R(T=10\,{\rm K})$ (with bias current $I=2\,\mu$A) of the different
bridges on each chip.
Due to the strong influence of the reduced temperature $t\equiv
T/T_c$ (for $T$ close to $T_c$) on the characteristic length scales,
i.e.~the London penetration depth $\lambda(t)$  and coherence length
$\xi(t)$, and on $I_c(t)$, the determination of $T_c$ plays an
important role for the comparison and interpretation of the
performance of pinning arrays with different $P_d$.
We defined $T_c$ by linear extrapolation of the $R(T)$ curves in the
transition region to $R=0$, i.~e.~$T_c$ marks the onset of
resistance.
For all samples we find $T_c\approx 8.5\,$K with a variation of a
few mK within each set of bridges, and $R_n=5.0\,\Omega$ to
$5.8\,\Omega$, depending on $n_p$ and $D$.
Within the sets ($D=const)$ with $n_p=const$ (scaled lattices),
$R_n$ varies by less than $\pm 2\,\%$ from bridge to bridge.
The plain film (from set \#4-A) has $R_n=5.0\,\Omega$, which yields a
normal resistivity $\rho_n=R_ndW/L=9.4\,\mu\Omega$cm.
With the relation $\rho\ell=3.72\times 10^{-6}\,\mu\Omega{\rm
cm}^2$ \cite{Mayadas72} we estimate for the mean free path $\ell= 4.0\,$nm.
%
All $I_c$ values were determined with a voltage criterion
$V_c=1\,\mu$V.


\begin{figure}[tbp]
\center{\includegraphics[width=8.5cm]{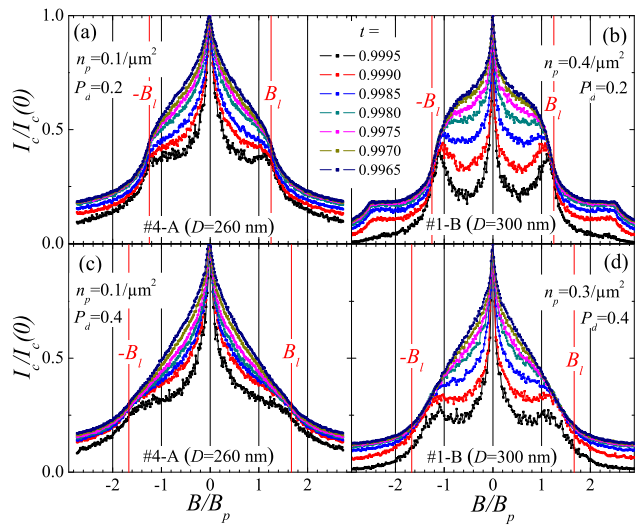}}
%
\caption{(color online)
$I_c/I_c(0)$ vs $B/B_p$ for randomly diluted arrays at different
reduced temperatures ($I_c$ increases with decreasing $t$).
Comparison of the dependence on the antidot density $n_p$ for samples
with $P_d=0.2$ (a),(b) and $P_d=0.4$ (c), (d).
\label{fig-latticedensity}}
\end{figure}

Fig.~\ref{fig-latticedensity} shows $I_c(B/B_p)$ patterns of four
randomly diluted bridges with similar antidot size at various
temperatures $t=T/T_c=0.9995\,\ldots\,0.9965$.
$I_c$ was normalized to its maximum value at $B=0$.
Two bridges (from set \#4-A) have the same antidot density
$n_p=0.1\,\mu\rm m^{-2}$ and dilution $P_d=0.2$ (a) and $P_d=0.4$
(c).
The two other bridges (from set \#1-B) have larger antidot density
$n_p=0.4\,\mu\rm m^{-2}$ with $P_d=0.2$ (b) and $n_p=0.3\,\mu\rm
m^{-2}$ with $P_d=0.4$ (d).

For the highest $t=0.9995$, a clear peak in $I_c(B)$ indicates
matching of the vortex lattice with the pinning array, as shown in
Fig.~\ref{fig-latticedensity}(b).
The position of the peak is located between $B_p$ and $B_l$.
This indicates that indeed ''pin density matching`` is observed;
however, as pointed out in \cite{Reichhardt07a}, for a diluted
periodic pinning array at $B=B_p$ the vortex configuration contains
numerous topological defects.
Hence, commensurability effects at $B_p$ can only be observed when
pinning is so strong, that the lattice distortion energy, associated
with the deviation from an ideal triangular vortex lattice, can be
overcome.
This is most likely to be observed for the samples with higher
density of pinning sites.
Accordingly, for a given $P_d$, the matching peak is more pronounced
in the samples with three and four times larger $n_p$.

With decreasing $n_p$ and $t$ and with increasing $P_d$ the peak in
$I_c(B)$ gradually transforms into a shoulder-like structure, located
close to $B_p$.
Following the evolution of $I_c(B)$ with decreasing temperature shows
that $I_c(B<B_p)/I_c(0)$ increases most whereas $I_c(B_l)/I_c(0)$ is
almost independent of $t$.
For the larger $P_d=0.4$, this leads gradually to a transformation of
the shoulder-like structure into a triangular-shaped $I_c(B)$ pattern
without any indication of matching effects, either at $B_p$ or $B_l$.
Nevertheless, $I_c$ for $P_d=0.4$ is still significantly enhanced
over $I_c$ for samples without antidots, as will be shown below.


In the following, we directly compare pinning arrays with different
dilution $P_d$ at the same reduced temperature.
%
%
%
%
%
\begin{figure}[tbp]
\center{\includegraphics[width=8.5cm]{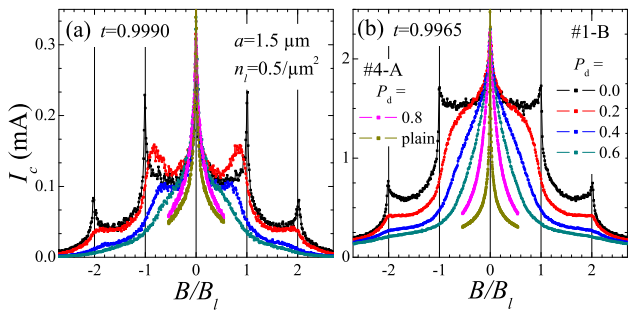}}
%
\caption{(color online)
$I_c$ vs $B/B_l$ at $t=0.9990$ (a) and 0.9965 (b) for arrays with fixed
lattice parameter $a$ and variable dilution $P_d$.
$I_c(B>B_l)$ decreases with increasing $P_d$.
\label{fig-transitiontoplain}}
\end{figure}
%
%
%
Fig.~\ref{fig-transitiontoplain} shows $I_c(B)$ patterns of samples
with fixed lattice parameter $a=1.5\,\mu$m ($n_l=0.5\,\mu{\rm
m}^{-2}$) at $t=0.9990$ (a) and $t=0.9965$ (b).
The $B$-axis is normalized to the lattice matching field $B_l$, which
is the same for all perforated bridges within this set.
The sample with $P_d=0$ shows pronounced peaks in $I_c(B)$ which are
located at $\pm B_l$ and $\pm 2B_l$, indicating a saturation number
$n_s\ge2$ \cite{Mkrtchyan72,Berdiyorov06b} for both temperatures.
In the samples with small dilution ($P_d= 0.2$ and 0.4) we also find
peaks in $I_c(B)$ for the higher temperature shown in
Fig.\ref{fig-transitiontoplain}(a).
The matching peaks are significantly broader than the matching peaks
of the undiluted bridge, and they are located at magnetic fields
between $B_l$ and $B_p(P_d)$.
This is also visible in Fig.~\ref{fig-latticedensity}(b) and (d).
With increasing $P_d$ we find a gradual transition of the $I_c(B)$
patterns at $B<B_l$ from the undiluted array ($P_d=0$) to the plain
film ($P_d=1)$.
Interestingly, for $t=0.9990$
[c.f.~Fig.\ref{fig-transitiontoplain}(a)] the diluted sample with
$P_d=0.2$ shows a higher $I_c$ than the undiluted sample ($P_d=0$)
for $B<B_l$, and very similar $I_c$ values for all other fields,
except for the matching fields $B_l$ and $2B_l$.
%


\begin{figure}[tb]
\center{\includegraphics[width=8.5cm]{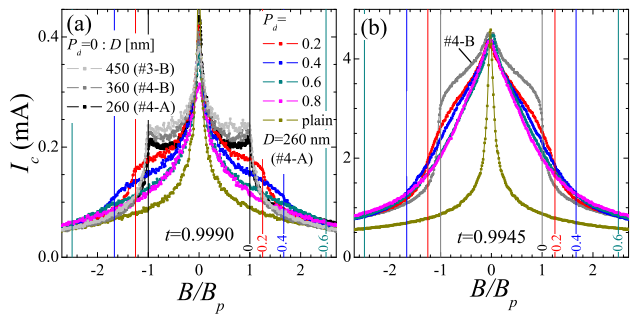}}
%
\caption{(color online)
$I_c$ vs $B/B_p$ at $t=0.9990$ (a) and 0.9945 (b)
for arrays with scaled lattice parameters
$a(P_d)$ (fixed antidot density $n_p=0.1\,\mu{\rm m}^{-2}$
and matching field $B_p$):
Comparison of samples with different dilutions $P_d$
and with different antidot diameters $D$ for $P_d=0$.
$I_c(B\lapprox B_p)$ decreases with decreasing $D$ (for $P_d=0$) and with increasing $P_d$.
Vertical solid lines indicate $\pm B_l(P_d)/B_p$, labeled with corresponding values for $P_d$.
\label{fig-IcBscaled}}
\end{figure}

Fig.~\ref{fig-IcBscaled} shows $I_c(B/B_p)$ patterns of bridges with
different $P_d=0\,\ldots\,1$ at $t=0.9990$ (a) and $t=0.9945$ (b).
In Fig.~\ref{fig-IcBscaled}(a) the data from set \#4-A ($D=260\,$nm)
are complemented by results from two undiluted bridges with
$D=360\,$nm (\#4-B) and $D=450\,$nm (\#3-B), in order to demonstrate
the effect of antidot size.
All bridges with $P_d=0$ and different $D$ show qualitatively the
same $I_c(B)$ patterns.
The major difference is a slight increase in $I_c$ at $B<B_p$ with
increasing $D$.
This dependence can be explained with the increase of the pinning
strength of the antidots with increasing $D$.
For $B<B_p$, each vortex can be captured by an antidot, and hence the
increasing pinning strength with $D$ leads to an increase of $I_c$.
In contrast, for $B>B_p$, vortices occupy interstitial pinning sites
where they are weakly pinned.
Hence, $I_c$ is drastically reduced and determined by the motion of
interstitial vortices, which should not depend on the antidot size.
This is exactly what we find experimentally, i.e.~$I_c(B>B_p)$ is
independent of $D$ for the samples with $P_d=0$.

With increasing $P_d$ (decreasing $a$ in order to keep the antidot
density $n_p$ constant) the shape of the $I_c(B/B_p)$ patterns is
strongly affected.
The diluted arrays show ''lattice matching`` effects around $B_l$.
It is interesting to note that with decreasing $t$ the matching
effects become less pronounced, and the $I_c(B)$ pattern approaches
more and more a triangular shape.
This shape is reminiscent of the $I_c(B)$ pattern of randomly
arranged antidots \cite{Kemmler06}.
This can be understood, as with increasing $P_d$ and fixed $n_p$, the
antidot arrangement approaches that of a random arrangement.
%


For $P_d=0$ from set \#4-A we have no data for the full range of $t$,
as the bridge was damaged after taking first data.
Hence, to facilitate the comparison of different $P_d$ at $t=0.9945$
[Fig.~\ref{fig-IcBscaled}(b)], we also show data from another bridge
with $P_d=0$ (from set \#4-B).
For both temperatures ($t=0.9990$ and $t=0.9945$) we find a decrease
in $I_c$ with increasing $P_d$ for $B<B_p$.
However, for $B>B_p$ the diluted arrays show an enhanced $I_c$ as
compared to the undiluted sample(s).
Both observations are in qualitative agreement with the simulations
in Ref.~\cite{Reichhardt07a}.
We do find that the enhancement of $I_c$ above $B_p$ persists up to
the highest dilution $P_d=0.8$.
This enhancement was explained in \cite{Reichhardt07a} with the
suppression of channeling of interstitial vortices for fields in the
range $B_l<B<B_p$.
For undiluted arrays, this channeling effect causes the rapid
decrease of $I_c$ with increasing $B$ slightly above $B_p$
[c.~f.~Fig.~\ref{fig-IcBscaled}].
Our experimental data clearly confirm that this rapid drop in $I_c$
at $B_p$ is absent for the diluted arrays.

\begin{figure}[tbp]
\center{\includegraphics[width=7.5cm]{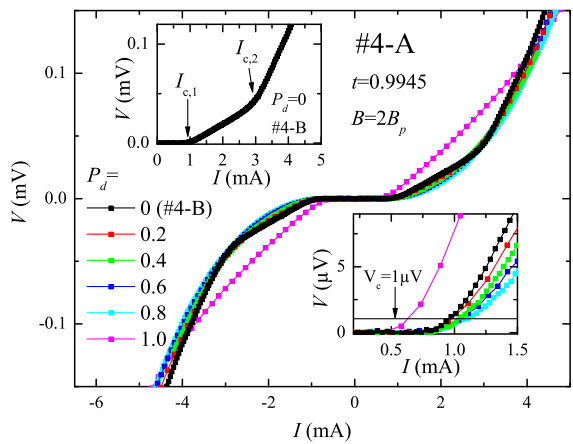}}
%
\caption{(color online)
$V(I)$ curves at $B=2B_p$ and $t=0.9945$ for arrays with scaled
lattice parameters $a(P_d)$ (fixed antidot density $n_p=0.1\,\mu{\rm m}^{-2}$
and matching field $B_p$).
Upper inset shows magnification of $V$ vs $I$ for $P_d=0$ with the
critical depinning current $I_{c,1}$($I_{c,2}$) of interstitial
(pinned) vortices.
Lower inset shows magnification of $V$ vs $I$ at small voltages;
$V$($I>$1\,mA) decreases with increasing $P_d$, except for the plain film ($P_d=1$) with lowest $I_c$.
\label{fig-IVscaled}}
\end{figure}

The suppression of channeling should also be visible in the
current-voltage-characteristics.
Fig.~\ref{fig-IVscaled} shows the $V(I)$-curves of differently
diluted samples at $B=2B_p$ and $t=0.9945$.
The $V(I)$-curves directly correspond to the $I_c(B)$ pattern shown
in Fig.~\ref{fig-IcBscaled}(b).
For the undiluted bridge at $B=2B_p$ one expects the same number of
vortices sitting in the antidots and at interstitial positions.
In this case we find two critical depinning currents
$I_{c,1}$,$I_{c,2}$ in the $V(I)$-curves [c.f.~upper left inset of
Fig.~\ref{fig-IVscaled}].
$I_{c,1}$ corresponds to the current above which a finite voltage
appears.
This voltage is caused by the motion of weakly pinned interstitials.
At a higher current $I=I_{c,2}$ the slope of the $IV$-curve changes.
This is due to the depinning of vortices sitting in the antidots.
The value of $I_{c,2}$ can also be found in the $I_c(B)$ pattern
shown in in Fig.~\ref{fig-IcBscaled} (b).
$I_{c,2}$ fits quite well to the critical current at the first
matching field $I_c(B_p)$.
All diluted samples ($P_d=0.2\ldots 0.8$) have a similar $I_{c,1}$
but do not show a pronounced slope change at $I_{c,2}$.
The lower right inset in Fig.~\ref{fig-IVscaled} clearly shows that
with increasing $P_d$ the voltage due to flux motion decreases.
I.~e., in the diluted pinning arrays dissipation is reduced, due to
the more effective suppression of vortex channeling.


In conclusion, we experimentally investigated Nb thin films with
triangular arrays of antidots, which have been randomly diluted, by
measurements of the critical current $I_c$ vs.~applied magnetic field
$B$ and current-voltage ($IV$) characteristics close to the
transition temperature $T_c$.
The antidot lattices could be tuned to find two different matching
effects, related to the antidot density and to the lattice parameter
of the antidot lattice, as predicted in \cite{Reichhardt07a}.
For samples with fixed lattice constant, with increasing dilution
$P_d$ a gradual transition from a periodic pinning array to a plain
film without pinning sites has been observed.
Obviously, with increasing $P_d$ the critical current decreases.
However, very close to $T_c$, for small dilutions ($P_d=0.2$) we do
find a broad peak in $I_c(B)$ located between $B_p$ and $B_l$,
corresponding to an increase in $I_c$ by removing 20\,\% of the
pinning sites.
We speculate that this counterintuitive effect is due to the
introduction of disorder; an understanding of this effect is still
lacking, and deserves further investigations.
On the other hand, for samples with fixed antidot density, an
increasing dilution corresponds to a gradual transition from a
periodic to purely random distribution of pinning sites.
Our experiments clearly show an enhancement of $I_c$ for magnetic
fields above $B_p$ with increasing $P_d$.
This was the main prediction in Ref.~\cite{Reichhardt07a} and can be
explained with the suppression of channeling of interstitial
vortices.
This effect is also observed in $IV$-measurements, i.e., the
suppression of channeling causes an increasing reduction in the
flux-flow voltage with increasing $P_d$.
As a consequence, the concept of introducing disorder by randomly
removing pinning sites in tailored periodic pinning arrays seems to
provide a feasible way for enhancing the critical current in
superconductors for magnetic fields above the matching field $B_p$.

This work was supported by the DFG via the SFB/TRR21 and in part by
the DFG Research Center of Functional Nanostructures.
M.~Kemmler gratefully acknowledges support from the
Carl-Zeiss-Stiftung.

\bibliography{dilutedPA}

\end{document}